\documentclass[twocolumn, usenatbib]{mnras}
\usepackage{savesym}
\usepackage{graphicx}
\usepackage{longtable}
\usepackage{changepage}

\expandafter\let\csname equation*\endcsname\relax
  \expandafter\let\csname endequation*\endcsname\relax 
\usepackage{subfig}
\usepackage{amsmath}
\usepackage{amssymb}
\usepackage{verbatim}
\usepackage[yyyymmdd,hhmmss]{datetime}
\usepackage{array}
\usepackage{times}
\usepackage{xcolor}

\DeclareRobustCommand{\DE}[3]{#2}
\let\DEthebibliography\thebibliography
\def\thebibliography{\DeclareRobustCommand{\DE}[3]{##3}\DEthebibliography}

\title[Magnetic fields and black hole evolution]{ Black hole–disc coevolution in the presence of magnetic fields
: \\ refining the Thorne limit with emission from within the plunging region   }
\author [Andrew Mummery]{Andrew Mummery$^1$\thanks{E-mail:
andrew.mummery@physics.ox.ac.uk} \\
$^1$Oxford Theoretical Physics, Beecroft Building,  Clarendon Laboratory, Parks Road, Oxford, OX1 3PU, United Kingdom 
}

\date{}

\begin{document}

\pagerange{\pageref{firstpage}--\pageref{lastpage}} \pubyear{2024}

\maketitle

\label{firstpage}

\begin{abstract} 
The accretion of material onto a black hole modifies the properties of that hole owing to the capture of both matter and radiation. Adding matter to the hole through an accretion disc generally acts to increase the  hole's spin parameter, while the capture of radiation generally provides a retarding torque. The balance between the torques provided by adding matter and radiation leads to a maximum spin parameter that can be obtained by a black hole which grows by accretion, known as the Thorne limit. In the simplest theory of thin disc accretion this Thorne limit has the value $a_{\bullet, {\rm lim}} \simeq 0.998$. The purpose of this paper is to highlight that any modification to theories of accretion flows also modify this limiting value, and to compute precisely the modification arising from a particular extension of accretion theory: the inclusion of bright emission from within the plunging region which is sourced from the magnetohydrodynamic stresses ubiquitously observed in simulations. {This extra emission further suppresses black hole spin-up and results in new, lower, limits on the final black hole spin.} {These limits} depend on the details of the magnetic stresses acting within the plunging region, but  typical values seen in simulations and observations would lower the limit to $a_{\bullet, {\rm lim}} \simeq 0.99$, a subtle but not negligible deviation. 
\end{abstract}

\noindent

\section{Introduction} 
The addition of material onto a black hole via an accretion flow will, provided a sufficiently large reservoir of material is available, profoundly modify the properties of the black hole itself. This is because the accreted matter carries both energy and angular momentum, the addition of which modifies the mass and spin parameters of the hole.  It was first argued by \cite{Bardeen1970} that an initially Schwarzschild black hole would be spun up to an extremal Kerr black hole (with dimensionless spin parameter $a_\bullet = 1$) once the mass parameter of the hole had changed from $M_\bullet = M_i$ to $M_\bullet = \sqrt{6}M_i$. In other words, roughly doubling the mass of the black hole by accretion also spins it up to maximal rotation. 

The \cite{Bardeen1970} argument only considers the effects of adding matter to the black hole, which is not the only form of stress energy present in an accreting system. \cite{Thorne1974} refined the analysis of \cite{Bardeen1970} by including the energy and angular momentum carried onto the black hole by the photon field emitted by the accretion flow. Photons carry angular momentum and energy, and therefore their capture also modifies the evolution of the black hole.  It is, generally speaking, difficult to capture a photon, even for a black hole. Photons which are counter-rotating with respect to the black hole's spin axis must ``work against'' the rotation of the hole, and have a larger capture cross section \citep{Thorne1974}. Therefore, photons generally provide a net {\it negative} angular momentum flux onto the black hole, spinning it down. \cite{Thorne1974} showed that incorporating the effects of photon capture results in a limiting black hole spin at which the torque from the accreting matter is balanced by the retarding torque of the radiation field, at which point the spin evolution of the black hole stops. This limiting spin has a very well known value $a_{\bullet, {\rm lim}} \simeq 0.998$, and is known as the ``Thorne limit''. 

What is perhaps less widely appreciated is that, while the spirit of the calculation is model-independent (photon capture will always act to spin-down the hole owing to the basic properties of their capture cross section), the actual numerical value of the limiting spin is a model dependent statement. Any modification to theoretical descriptions of accretion flows will also modify this limiting value, and therefore this limit should be re-examined as our theories of accretion flows develop from their 1970s forms. 

One modification discussed in the original \cite{Thorne1974} analysis is the potential implications of magnetic {fields} acting in the innermost regions of the disc. Indeed, \cite{Thorne1974} notes that magnetic {fields} could modify the vanishing ISCO (innermost stable circular orbit) stress assumption employed in their analysis, which would result in an increased photon flux being emitted on small scales (photons which are therefore most likely to be captured), increasing the net spin-down effect of the radiation field. We now of course know that magnetic stresses are essential for driving the accretion process \citep{BalbusHawley91}, and flux-freezing results in large enough magnetic stresses in the plunging region to keep the accretion flow hot in the innermost regions \cite{Krolik99, Gammie99}. This basic result (which cannot be captured by the $\alpha$-viscosity toy model, which is unphysical in this regime) has been confirmed by numerous General Relativistic Magnetohydrodynamic (GRMHD) simulations over the years \citep[e.g.,][]{Noble10, Schnittman16, Wielgus22}. Some of the implications for black hole spin evolution in the presence of a magnetic ISCO stress were first discussed in \cite{AgolKrolik00}, although their analysis neglected photons emitted from the plasma on the plunge itself. Indeed, we do not believe any calculation including the captured photon flux emitted from plunging gas has been performed in the literature.

It is the purpose of this short paper to revisit the \cite{Thorne1974} limit calculation using extended models for thin accretion flows which now include emission from within the ISCO. We shall show that the additional dissipation at and within the ISCO results in an increased retarding radiation torque, and therefore a slower rate of black hole spin-up and lower equilibrium spin limits.

{This is an important result, as the study of co-evolving black hole–disc systems have implications for the present day properties of the population of supermassive black holes which reside in galactic centers and, ultimately, of the properties of their galaxies themselves \citep{Silk98}. Unlike accreting stellar mass black hole systems, which generally do not have a large enough mass budget available to accrete and change the hole's properties, supermassive black holes are expected to change their natal mass by large factors across cosmic history (by accreting circumnuclear material, and/or undergoing mergers), and therefore will have spin parameters set by the particular history of their cosmic evolution \citep[see e.g.,][for a recent discussion]{Piotrowska24}. A reduction in the rate of black hole spin-up caused by accretion, and lower saturating spin values, therefore must be taken into account when inferring evolution histories from present day spin measurements. As the spin of a black hole sets (to leading order) the radiative efficiency of the accretion process and the power of relativistic jets launched from the disc \citep[if powered by the][mechanism]{Blandford77}, this black hole spin evolution also influences all of the feedback mechanisms accreting black holes have on galaxy evolution \citep{Fabian12b} and, therefore, on cosmological galaxy evolution.      } 

The layout of the paper is as follows. In section \ref{formalism} we lay out the \cite{Thorne1974} formalism, and extend it to include photons emitted from fluid elements which are not on circular orbits. In section \ref{results} the results of this analysis are presented, and are discussed in section \ref{conclusions}. A photon capture algorithm which extends the \cite{Thorne1974} procedure to include emitting gas with a non-zero radial velocity is presented in Appendix \ref{algorithm}. 

\section{Formalism}\label{formalism}
\subsection{The metric}
For the remainder of this paper we shall work in a units system in which $G = 1 = c$.  The Kerr metric in Boyer-Lindquist coordinates takes the following form, which we  present here in terms of its invariant line element  ${\rm d}s^2 \equiv g_{\mu\nu} {\rm d}x^\mu {\rm d}x^\nu$
\begin{multline}\label{metric}
    {\rm d}s^2 = - \left(1 - {2 M_\bullet r \over \Sigma }\right)  {\rm d}t^2 - {4 M_\bullet r a \sin^2 \theta \over \Sigma} \, {\rm d}t \, {\rm d}\phi  + {\Delta \over \Sigma }\, {\rm d}r^2 \\ + \Sigma \, {\rm d}\theta^2   + \left(r^2 + a^2 + {2 a^2 r \sin^2 \theta \over \Sigma } \right)\sin^2 \theta \, {\rm d}\phi^2 ,
\end{multline}
where $M_\bullet$ is the black hole mass, and $a$ is the angular momentum constant of the black hole $a = J/M_\bullet$ (with the same dimensions as the mass in this unit system), where $J$ is the total angular momentum of the black hole. We have defined the following shorthand functions following the notation of \cite{Bardeen72}
\begin{align}
    \Delta &= r^2 - 2M_\bullet r + a^2, \\ 
    \Sigma &= r^2 + a^2 \cos^2 \theta . 
\end{align}
 We shall also work with the dimensionless black hole spin $a_\bullet$ in this paper 
\begin{equation}
    a_\bullet \equiv a/M_\bullet, \quad |a_\bullet| < 1. 
\end{equation}
The coordinates have the following physical interpretation, $t$ is the time as measured at infinity, and the three spatial coordinates $(r, \theta, \phi)$ which have their usual quasi-spherical meaning. 

\subsection{Black hole evolution driven by accretion}
Consider a fluid element of rest mass $\delta m_0$ which is accreted onto a black hole through a disc. After it crosses the event horizon and falls down to the singularity, it will increase the angular momentum of the black hole by 
\begin{equation}\label{nj}
    \delta J = \delta m_0\,  U_\phi(r_+),
\end{equation}
and will increase the mass parameter of the black hole by 
\begin{equation}\label{nm}
    \delta M_\bullet = - \delta m_0 \, U_0(r_+),
\end{equation}
where $U_\phi$ and $U_0$ are the azimuthal and time components of the fluid element's covariant four-velocity, evaluated at the point the fluid element crosses the event horizon $(r_+/M_\bullet = 1 + \sqrt{1 - a_\bullet ^2})$. These coupled evolutionary equations (\ref{nj} and \ref{nm}) were first solved by \cite{Bardeen1970} and – under the assumption that $U_\phi$ and $U_0$ at the horizon are given by their values at the last stable circular orbit – can be used to show that thin disc accretion inevitably leads to $a_\bullet = 1$ after a finite amount of mass is added to the black hole. This argument, however, neglects the impact of the radiation emitted via the accretion process on the black hole's evolution, as was first pointed out by \cite{Thorne1974}. 

A fluid element which spirals onto the black hole through a radiatively efficient accretion flow will also emit radiation over the course of its journey through the disc. Some of this radiation will be captured by the black hole, and as photons carry energy and angular momentum this photon flux will impact the evolution of the black hole's mass and spin parameters. The evolutionary equations for the black hole are thus modified to 
\begin{equation}\label{ej}
    \delta J = \delta m_0 \, U_\phi(r_+) + \delta J_{\rm rad},
\end{equation}
and
\begin{equation}\label{em}
    \delta M_\bullet = - \delta m_0 \, U_0(r_+) + \delta E_{\rm rad},
\end{equation}
where $\delta J_{\rm rad}$ is the total angular momentum flux onto the black hole from all of the photons which were emitted by the fluid element over its journey through the disc, and  $\delta E_{\rm rad}$ the corresponding total photon energy flux. Combining these modified evolutionary equations, one can compute the evolution of 
\begin{equation}
    {{\rm d} a_\bullet \over {\rm d}\ln M_\bullet} = M_\bullet {{\rm d}m_0\over {\rm d}M_\bullet} {{\rm d} \over {\rm d}m_0}\left({J\over M_\bullet^2}\right) , 
\end{equation}
or explicitly 
\begin{equation}\label{fund}
    {{\rm d} a_\bullet \over {\rm d}\ln M_\bullet} = {1\over M_\bullet}\left[ {U_\phi(r_+) + {\rm d} J_{\rm rad}/{\rm d}m_0 \over -U_0(r_+) + {\rm d} E_{\rm rad}/{\rm d}m_0}\right] - 2a_\bullet ,
\end{equation}
which is the most convenient form to use analytically. Clearly, the non-trivial element of this calculation lies in computing the photon flux quantities ${\rm d} J_{\rm rad}/{\rm d}m_0$ and ${\rm d} E_{\rm rad}/{\rm d}m_0$. The technique for calculating these quantities was described by \cite{Thorne1974}, for which the interested reader may wish to turn for full details.  For our purposes it is simplest to write down the final result of this analysis, before examining the physical quantities involved and seeing why this must be the correct result.  Over the course of the fluid elements evolution through the disc the angular momentum flux onto the black hole from its emitted photon field is given by
\begin{multline}
    {{\rm d} J_{\rm rad} \over {\rm d}m_0} = {1\over \dot M} \int_{r_+}^\infty \int_0^{\pi/2} \int_0^{2\pi} \bigg\{ n_\phi(r, \Theta, \Phi) C(r, \Theta, \Phi)  \\ S(\Theta, \Phi) \cos \Theta \sin\Theta \,  4 \pi r F(r) \bigg\} 
    \, {\rm d} \Phi \, {\rm d}\Theta \, {\rm d}r ,
\end{multline}
while the energy flux is similarly given by
\begin{multline}
    {{\rm d} E_{\rm rad} \over {\rm d}m_0} = {1\over \dot M} \int_{r_+}^\infty \int_0^{\pi/2} \int_0^{2\pi} \bigg\{ (- n_0(r, \Theta, \Phi))\,  C(r, \Theta, \Phi)  \\ S(\Theta, \Phi) \cos \Theta \sin\Theta \,  4 \pi r F(r) \bigg\}
    \, {\rm d} \Phi \, {\rm d}\Theta \, {\rm d}r . 
\end{multline}
Let us begin by defining the quantities involved. The angles $\Theta$ and $\Phi$ represent the emission angles of each photon {\it in the rest frame} of the orbiting fluid element, where $\Theta$ is the angle down from the vertical (defined by the disc axis, which we shall assume is also the black hole spin axis) and $\Phi$ the angle around the disc axis. The function $S(\Theta, \Phi)$ is an emissivity shape function, which is formally arbitrary and describes the angles into which photons are emitted in the rest frame (e.g., isotropically, from an electron-scattering atmosphere, etc.). This shape function must only satisfy the normalisation condition 
\begin{equation}
    \int_0^{2\pi} \int_0^{\pi/2}   S(\Theta, \Phi) \cos \Theta \sin\Theta  \, {\rm d}\Theta \, {\rm d} \Phi = 1. 
\end{equation}
The factor $\sin \Theta$ here is the usual Jacobian factor of spherical coordinates, while the factor $\cos\Theta$ comes into the analysis as the disc flux escapes in the vertical direction, and so it is the vertical component of the photon stress-energy tensor which we are ultimately therefore interested in. 

The quantity $F(r)$ represent the total locally emitted flux of the disc at each radius, meaning that $2\pi r F(r) \, {\rm d}r$ is the locally liberated luminosity from the surface of each disc annulus (the extra factor 2 in the above expressions therefore making this the total luminosity emitted from each disc annulus accounting for both of the upper and lower disc faces). This means that the integral 
\begin{equation}
     \int_{r_+}^\infty 4 \pi r F(r) \, {\rm d}r = L_{\rm bol}, 
\end{equation}
where $L_{\rm bol}$ is the bolometric luminosity of the disc. 

The final, and most important, factors are $C(r, \Theta, \Phi)$ and the components $n_\mu(r, \Theta, \Phi)$. The function $C(r, \Theta, \Phi)$ is a ``capture function'', and is equal to $C=1$ if the photon emitted from radius $r$ at rest-frame angles $\Theta, \Phi$ is ultimately captured by the black hole, and $C=0$ otherwise. The components $n_\mu(r, \Theta, \Phi)$ represent the normalised covariant four-velocity of the photon emitted from radius $r$ at rest-frame angles $\Theta, \Phi$ in the Boyer-Lindquist coordinate frame. Therefore $n_\phi$ represent the normalised angular momentum of the angular momentum carried by the photon, and $-n_0$ the energy. This normalisation condition is chosen so that $n^0=1$ in the rest frame of the fluid element, as the total energy emitted from all the photons is already fixed by the inclusion of the factor $4\pi r F(r)$. 

Therefore, if $C=1$ for all photons, the above integrals would simply count the total energy and angular momentum carried by all photons emitted from the disc per unit time. By normalising both integrals by $1/\dot M$, where $\dot M$ is the physical mass accretion rate in the disc, this then becomes the angular momentum and energy carried by the photon field per unit rest mass accreted. The inclusion of the capture function $C$ therefore specialises these integrals to only those photons which are captured by the black hole, and therefore they represent the angular momentum and energy flux onto the black hole from the disc radiation field, per unit accreted mass. 

As a general rule, it is more likely that a photon with angular momentum component pointing in the opposite direction to the black hole's angular momentum will be captured by the black hole. In some sense these photons have to ``work against'' the dragging of spacetime in the direction of the black hole's spin. Therefore, it is generally the case that more negative angular momentum photons are captured by the black hole, and the effect of disc radiation is ultimately to produce a retarding torque on the black hole, stopping it from reaching the \cite{Bardeen1970} result of $a_\bullet = 1$. 

The degree to which the disc photon field acts to retard the spin up of the black hole is dependent on the precise model for the accretion flow itself. This model dependence enters in two distinct physical areas. Firstly, the spin up itself is dependent on the components of the fluid energy and angular momentum at the horizon, and therefore is sensitive to the assumed dynamics of the accretion flow. Secondly, the locally emitted flux $F(r)$ naturally sets the overall scale at which photons can/cannot effect this spin up, and so the spin evolution is sensitive to the assumed thermodynamics of the disc flow. 

\cite{Thorne1974} solved the evolutionary equation (\ref{fund}) under the assumption that the orbital fluid components $U_\phi(r_+), U_0(r_+)$ where given by the values associated with the circular motion at the innermost stable circular orbit (ISCO), where it was then assumed they dropped out of the flow without communicating with the disc further. Correspondingly, the disc flux $F(r)$ was assumed to vanish at the ISCO under the classical ``vanishing ISCO stress'' assumption. \cite{Thorne1974} noted that in the presence of magnetic fields \citep[which we now know are essential to the accretion process][]{BalbusHawley91}, this dynamical assumption may be modified, something we now know is true from numerous GRMHD simulations \citep[see e.g.,][among many others]{Noble10, Schnittman16, Wielgus22}. \cite{Thorne1974} also noted that this modification to the dynamics would only have a very small effect on the ultimate spin up of the black hole, even for moderate $\sim 10\%$ deviations in $U_\phi(r_+)$ and $U_0(r_+)$ from their ISCO values. 

This part of the \cite{Thorne1974} argument is correct (and will be verified in this paper), however what was not noted at that time was that a modification to $U_\phi(r_+)$ and $U_0(r_+)$ will result in a corresponding increase in the disc dissipation, which can substantially modify $F(r)$ in the innermost disc regions. We will show in this paper that this modification to $F(r)$ is much more important than the effects of a pure change in the disc dynamics, and results in further, more substantial, modification of the limiting spin up value. 

Before we compute these final spin-up values, we must first construct the capture function $C(r, \Theta, \Phi)$ and relate the photons covariant four velocity $n_\lambda$ to the rest-frame emission angles $\Theta$ and $\Phi$. 

\subsection{Locally inertial frames in the Kerr metric}
The Kerr spacetime can, at all radii $r > 0$, always be described by (an infinite family of) locally inertial coordinate systems, in which the  metric is locally flat. This infinite family of such coordinate systems at each radius are simply related by (formally arbitrary) rotations and Lorentz boosts. As the construction of a rotation/Lorentz transformation between any two inertial frames is trivial, the key ingredient in this analysis is finding any sufficiently general local coordinate system to work with as a base transformation. 

Such a coordinate system was first identified by \cite{Bardeen72}, and is known as the Zero Angular Momentum Observer (hereafter ZAMO) frame\footnote{Note that \citealt{Bardeen72} originally named this the Locally Non Rotating Frame. Both names remain in use in the literature.}. The transformation from the Boyer-Lindquist coordinate system to the ZAMO frame is described by the following matrix
\begin{equation}
    E^{(a)}_\mu = 
    \begin{pmatrix}
        \sqrt{\Sigma \Delta \over A} & 0 & 0 & 0 \\
        0 & \sqrt{\Sigma \over \Delta} & 0 & 0 \\
        0 & 0 & \sqrt{\Sigma} & 0 \\
        - {2 {aM_\bullet r} \over \sqrt{\Sigma A}} & 0 & 0 & \sqrt{A  \over \Sigma \sin^2\theta }  
    \end{pmatrix} ,
\end{equation}
Note this transformation is well defined only for $r > r_+$. The corresponding inverse transformation is defined by the identity $E^{(a)}_\mu E_{(b)}^\mu = \delta^a_b$, and is given by 
\begin{equation}
    E_{(a)}^\mu = 
    \begin{pmatrix}
        \sqrt{A \over \Sigma \Delta} & 0 & 0 & 0 \\
        0 & \sqrt{\Delta \over \Sigma} & 0 & 0 \\
        0 & 0 & \sqrt{1 \over {\Sigma}} & 0 \\
         {2 {a M_\bullet r \sin \theta} \over  \sqrt{\Sigma A \Delta }} & 0 & 0 & \sqrt{\Sigma \sin^2\theta \over A} 
    \end{pmatrix} .
\end{equation}
Here we define 
\begin{equation}
    A = (r^2 + a^2)^2 - a^2 (r^2 - 2M_\bullet r + a^2) \sin^2 \theta ,
\end{equation}
again following the notation of \cite{Bardeen72}. To be clear in describing which quantities are evaluated in which frames, we use the notation $X^{(a)}$ (i.e., Roman alphabet and bracketed index) for a quantity evaluated in the ZAMO frame, while we use $X^\mu$ (i.e., Greek alphabet and non-bracketed index) for the Boyer-Lindquist coordinate system. 

This transformation  corresponds physically to describing physical quantities by their components as measured in the local ZAMO observer’s frame. For example, the four-velocity a ZAMO observer would measure is $U^{(a)} = E^{(a)}_{\mu} U^\mu$, where $U^\mu$ is the four-velocity measured in Boyer-Lindquist coordinates. One can simply verify that 
\begin{equation}
     g_{(a)(b)} = E^{\mu}_{(a)} E^{\nu}_{(b)} g_{\mu\nu} = \eta_{(a)(b)} = {\rm diag}(-1, 1, 1, 1) ,
\end{equation}
meaning that one can use the transformation laws of special relativity in this frame. 

\subsection{From the fluid rest frame to Boyer-Lindquist coordinates}
Assume that a fluid element moving with four-velocity $U^\mu$ (in Boyer-Lindquist coordinates), emits a photon which, {\it in the fluid rest frame}, moves in the direction $\widetilde n^{(\alpha)} \equiv \widetilde p^{\, (\alpha)} / \widetilde p^{\, (t)}$ (here $\widetilde p^{\, (\alpha)}$ is the photon four-velocity). We once again take care to distinguish the different frames of reference as far as possible, with quantities evaluated in the fluid rest frame being denoted $\widetilde X^{(\alpha)}$ (i.e., with a tilde, the Greek alphabet and bracketed indices). For the purposes of our calculation it will be important to determine what this rest-frame direction corresponds to in the Boyer-Lindquist coordinate system.

This direction four-vector can be defined in terms of two emission angles in the rest frame of the fluid element 
\begin{equation}
    \widetilde n^{(\alpha)} = 
    \begin{pmatrix}
        1 \\
        \sin \Theta \cos \Phi \\
        \sin \Theta \sin \Phi \\
        \cos \Theta 
    \end{pmatrix},  
\end{equation}
where we use the notation of \cite{Thorne1974} to distinguish these rest frame emission angles from the metric coordinates $\theta, \phi$. 

The Lorentz transform from the rest frame of the fluid to the ZAMO frame is given by the standard 4x4 special relativistic transformation law 
\begin{equation}
\Lambda^{(a)}_{(\alpha)} \equiv 
    \begin{pmatrix}
    \gamma & +\gamma \vec \beta^{\, T} \\
    +\gamma \vec \beta & \mathbb{I}_3 + (\gamma - 1) {\vec \beta \vec \beta^{\, T} / \beta^2} & 
    \end{pmatrix} ,
\end{equation}
where the four-velocity of the fluid element in the ZAMO frame defines $\gamma$ and $\vec \beta$ 
\begin{equation}
    U^{(a)} = E^{(a)}_{\mu} U^\mu \equiv \gamma 
    \begin{pmatrix}
        1 \\  \vec \beta
    \end{pmatrix} ,
\end{equation}
the 3x3 identity matrix is denoted $\mathbb{I}_3 = {\rm diag}(1, 1, 1)$, the transpose operation is denoted with a superscript $T$, and $\beta^2 \equiv \delta_{i j}\beta^i \beta^j$. 

Combining these results one can relate the emission direction in the fluid rest frame to that observed in the coordinate frame 
\begin{equation}\label{nbl}
    n^\mu = E^\mu_{(a)} \Lambda^{(a)}_{(\alpha)} \widetilde n ^{(\alpha)} ,
\end{equation}
with corresponding covariant four-vector $n_\lambda = g_{\lambda\mu} n^\mu$. This covariant form will be particularly important as it describes the (normalised) angular momentum and energy of the photon in Boyer-Lindquist coordinates, which are the quantities which effect the evolution of the hole. With $n_\lambda$ determined, the capture function $C(r, \Theta, \Phi)$ can be determined with a modified version of the original \cite{Thorne1974} algorithm, presented in Appendix \ref{algorithm}. This modification is required to account for those photons emitted from fluid elements which are on plunging orbits, which were not originally considered by \cite{Thorne1974}. 

As noted previously \citep[e.g., in][]{Bardeen72, Thorne1974}, this result (eq. \ref{nbl}) can be written in a particularly simple final form for fluid elements following an equatorial circular orbit.  For plunging fluid elements a similar result can be derived, but it is not particularly illuminating. 

\subsection{The full formalism}
We are now in a position to construct our full calculation of the limiting black hole spin. We shall assume, for the purposes of calculating the photon capture function, that the disc fluid follows circular orbits outside of the ISCO, and upon crossing the ISCO plunges toward the black hole on a trajectory well approximated by a geodesic infall. This results in four-velocity components  \citep[e.g.,][]{MTW}
\begin{align}
U^r &= 0,  \\
U^\phi &=  \frac{\sqrt{{M_\bullet}/{r^3}}}    {\left( 1 - {3M_\bullet}/{r} + 2a \sqrt{{M_\bullet}/{r^3} } \right)^{1/2}}, \\
U^0 &= \frac{1+a\sqrt{{M_\bullet}/{r^3}}}{\left({1 - {3M_\bullet}/{r} + 2a\sqrt{{M_\bullet}/{r^3} } }\right)^{1/2}} ,
\end{align}
for radii larger than the ISCO. Within the ISCO the fluid is assumed to spiral inwards while conserving (to leading order) its angular momentum and energy, which results in \citep{MummeryBalbus22PRL} 
\begin{align}
U^r &= - \sqrt{2M_\bullet \over 3 r_I} \left( {r_I \over r} - 1\right)^{3/2} , \label{ur} \\
U^\phi &= {2M_\bullet \gamma_I a  + J_I(r -2M_\bullet ) \over r(r^2 - 2M_\bullet r + a^2)} , \\
U^0 &=  {\gamma_I (r^3 + ra^2 + 2M_\bullet a^2) - 2J_I M_\bullet a \over r(r^2 - 2M_\bullet r + a^2)} ,
\end{align}
where we denote by $r_I$ the ISCO radius, and 
\begin{align}
J_I &= 2\sqrt{3} M_\bullet \left( 1 - {2a\over 3 \sqrt{M_\bullet  r_I}}\right), \\
\gamma_I &= \sqrt{1 - {2M_\bullet\over 3 r_I}}.
\end{align}
These four-velocity components are used for the purposes of calculating the photon emission properties and capture function. These fluid velocities are not exact solutions of the fluid equations, as angular momentum and energy are not perfectly advected with the flow. There are deviations at the percent level (this is after all the origin of the ISCO stress). These discrepancies do not substantially alter the photon capture physics. 

The locally emitted disc flux is given by the extended disc model of \cite{Mummery24Plunge}, which includes emission from within the ISCO of the black hole's spacetime. The key component which enters the spin-down expressions is the locally radiated flux, which outside of the ISCO is given by 
\begin{multline}\label{Fr_out}
F(r) = {3 \dot M  \over 8 \pi M_\bullet^2} {f_2(x) \over x^6} \Bigg[ f_1(x) - \left({x_I \over x}\right) f_1(x_I) (1 - \delta_{\cal J}) \Bigg]  , 
\end{multline}
where 
\begin{align}
    x &\equiv \sqrt{r \over M_\bullet}, \quad x_I \equiv \sqrt{r_I \over M_\bullet}, \\ 
    f_1(x) &=1- {3a_\bullet \over 2 x} \ln(x) + {1\over x} \sum_{\lambda = 0}^{2} k_\lambda \ln\left|x - x_\lambda\right|, \\
    k_\lambda &\equiv {2 x_\lambda - a_\bullet(1 + x_\lambda^2)  \over 2(1 - x_\lambda^2)} , \\
    x_\lambda &= 2 \cos\left[ {1\over 3} \cos^{-1}(-a_\bullet) - {2\pi\lambda\over3}\right]  , \\
    f_2(x)&= \left[ 1 -{ 3 \over  x^{2} } + { 2 a_\bullet \over x^{3}} \right]^{-1} .
\end{align}
This is precisely the generalisation of the classical \cite{NovikovThorne73, PageThorne74} relativistic thin disc solutions, in the presence of a finite ISCO stress. This finite ISCO stress is parameterised by $\delta_{\cal J}$, which can be thought of as the fraction of the specific angular momentum `passed back' into the disc from the $r \leq r_I$ region by whatever process is generating the stress at the ISCO. In the limit $\delta_{\cal J} \to 0$, this flux returns to the model used in \cite{Thorne1974}. 

Within the ISCO we use the solutions of the intra-ISCO thermodynamic equations derived in \cite{MummeryBalbus2023}. We assume that the flow is radiation pressure dominated (as would be the case for a black hole disc accreting at high accretion rates), so that the intra-ISCO radiative flux is 
\begin{equation}\label{FRin}
    F(r<r_I) = F_I \left({r\over r_I}\right)^{-17/7} \left[{1\over \epsilon} \left({r_I \over r} - 1\right)^{3/2} + 1\right]^{-4/28} ,
\end{equation}
where $\epsilon \simeq 10^{-3}$ is equal to $\epsilon \equiv  \sqrt{3r_Ic_s^2/2M_\bullet}$ where $c_s$ is the speed of sound in the inner disc regions  \citep[normalised by the speed of light in our unit system; see][for more details]{MummeryBalbus2023}. In this expression $F_I$ is equal to the extra-ISCO expression (equation \ref{Fr_out}) evaluated at $r=r_I$. 

The final element of our analysis is to determine the fluid elements angular momentum and energy as they cross the event horizon. While the precise modifications to the fluid behaviour  within the ISCO will depend on the details of the angular momentum transport by magnetic stresses, it is possible to relate the fluid angular momentum at the horizon to the ISCO stress boundary condition discussed above. For there to be a finite ISCO stress the fluid's event horizon angular momentum must be reduced by a factor $\delta_{\cal J}$ from its ISCO angular momentum\footnote{Formally this ignores the possibility that the ISCO stress could be sourced from an external torque, such as that driven by magnetic field lines threading a spinning black hole's horizon and the inner disc. In this case the accretion flow would be directly tapping the spin energy of the black hole, which could lead to substantially lowered spins \cite{AgolKrolik00}.}, i.e.,  
\begin{equation}
    U_\phi(r_+) = J_{\rm horizon} = J_I (1 - \delta_{\cal  J}) . 
\end{equation}
This reduced angular momentum will also modify the energy of the fluid elements. {Kerr metric circular orbits satisfy the energy-angular momentum relationship \citep[e.g.,][]{PageThorne74} }
\begin{equation}
    U_0' = - \Omega U_\phi ' , \quad \Omega = {\sqrt{M_\bullet/r^3} \over 1 + a\sqrt{M_\bullet/r^3}} ,
\end{equation}
where a prime denotes a radial gradient. While the fluid motion will not be precisely that of orbits with constants of motion of circular motion (there remains a stress after all), we can use this result to estimate the fluid's (dimensionless) energy parameter at the horizon, by assuming that the angular momentum deviation is small and the orbital elements will therefore be close to their circular values. This implies
\begin{equation}
    -U_0(r_+) = \gamma_{\rm horizon} = \gamma_I  - \Omega_I \delta_{\cal  J} J_I, 
\end{equation}
where $\Omega_I$ is the angular velocity of the disc material at the ISCO.

We therefore now have all of the ingredients required to compute the evolution of black hole properties resulting from the accretion of material through a thin disc which contains a finite ISCO stress and emission from within the plunging region. 

\section{Results}\label{results}
\subsection{The captured fraction}
\begin{figure}
    \centering
    \includegraphics[width=0.95\linewidth]{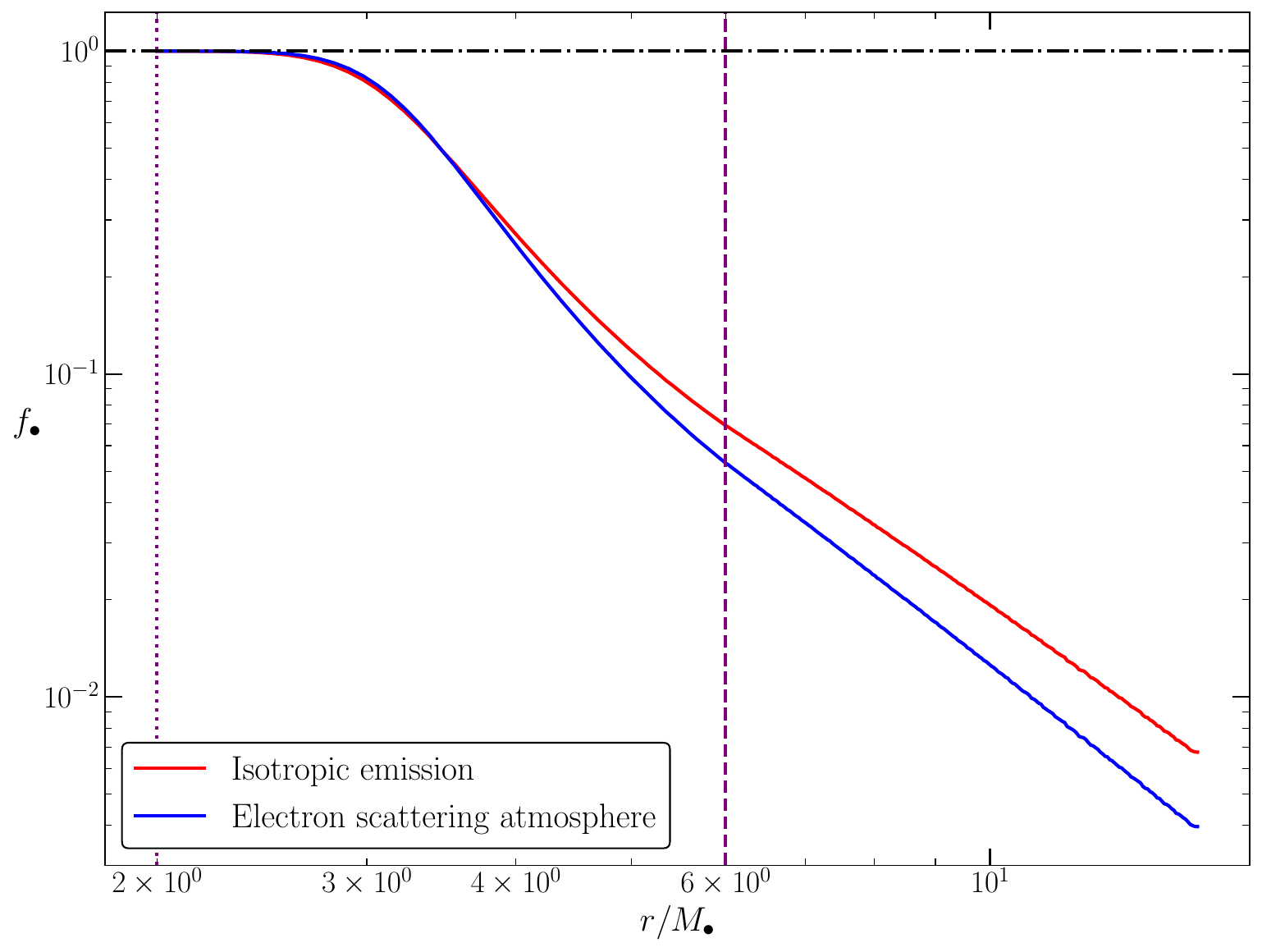}
    \includegraphics[width=0.95\linewidth]{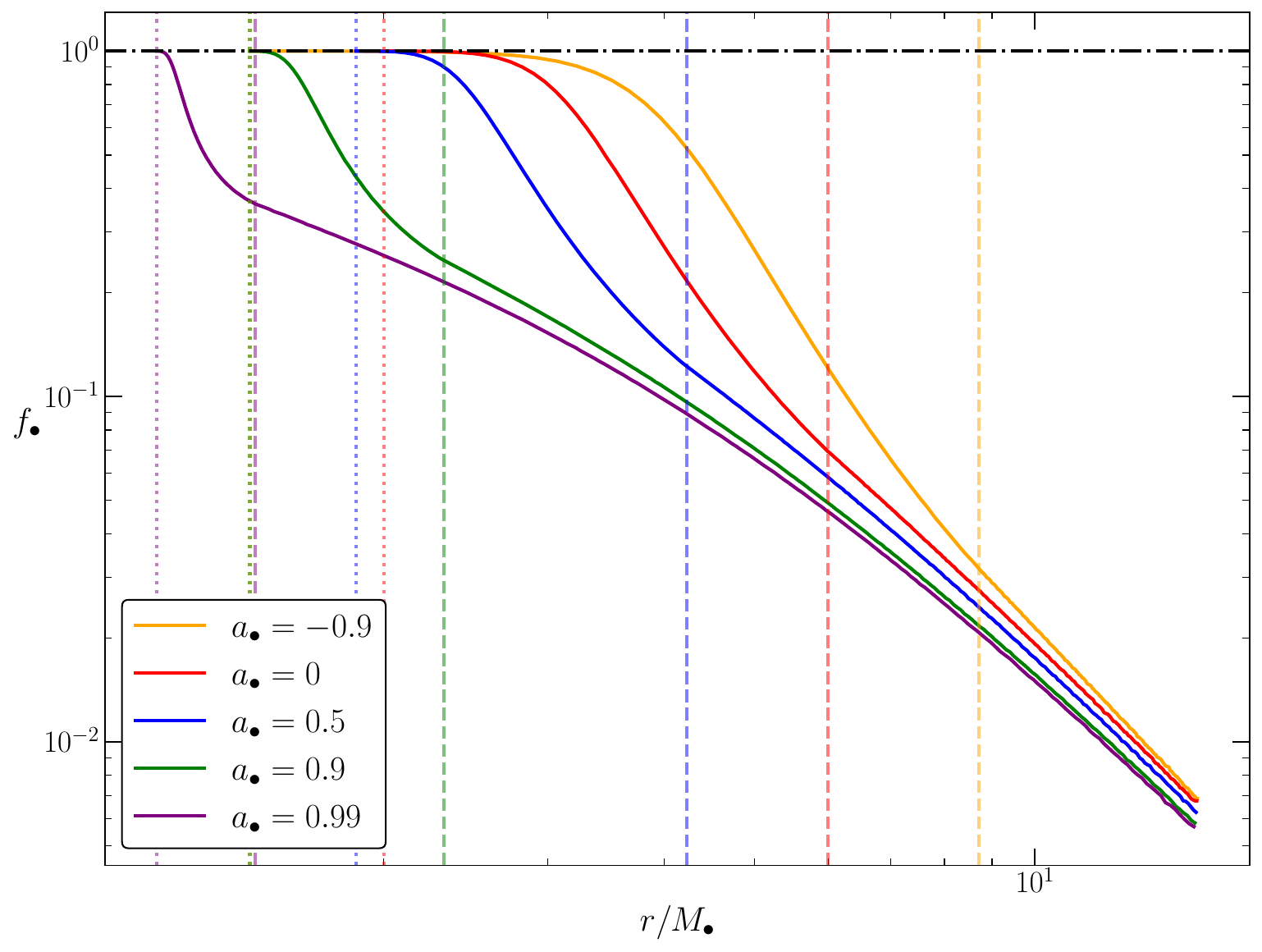}
    \caption{The fraction of emitted photons which are ultimately captured by the black hole $(f_\bullet)$ plotted as a function of disc radius for different rest-frame emission laws in the Schwarzschild metric (top), and for different black hole spins and isotropic rest-frame emission (bottom). For each black hole spin the ISCO radius is denoted by vertical dashed lines, while the horizon radius is denoted by vertical dotted lines.  }
    \label{fig:capfac}
\end{figure}

\begin{figure*}
    \centering
    \includegraphics[width=0.49\linewidth]{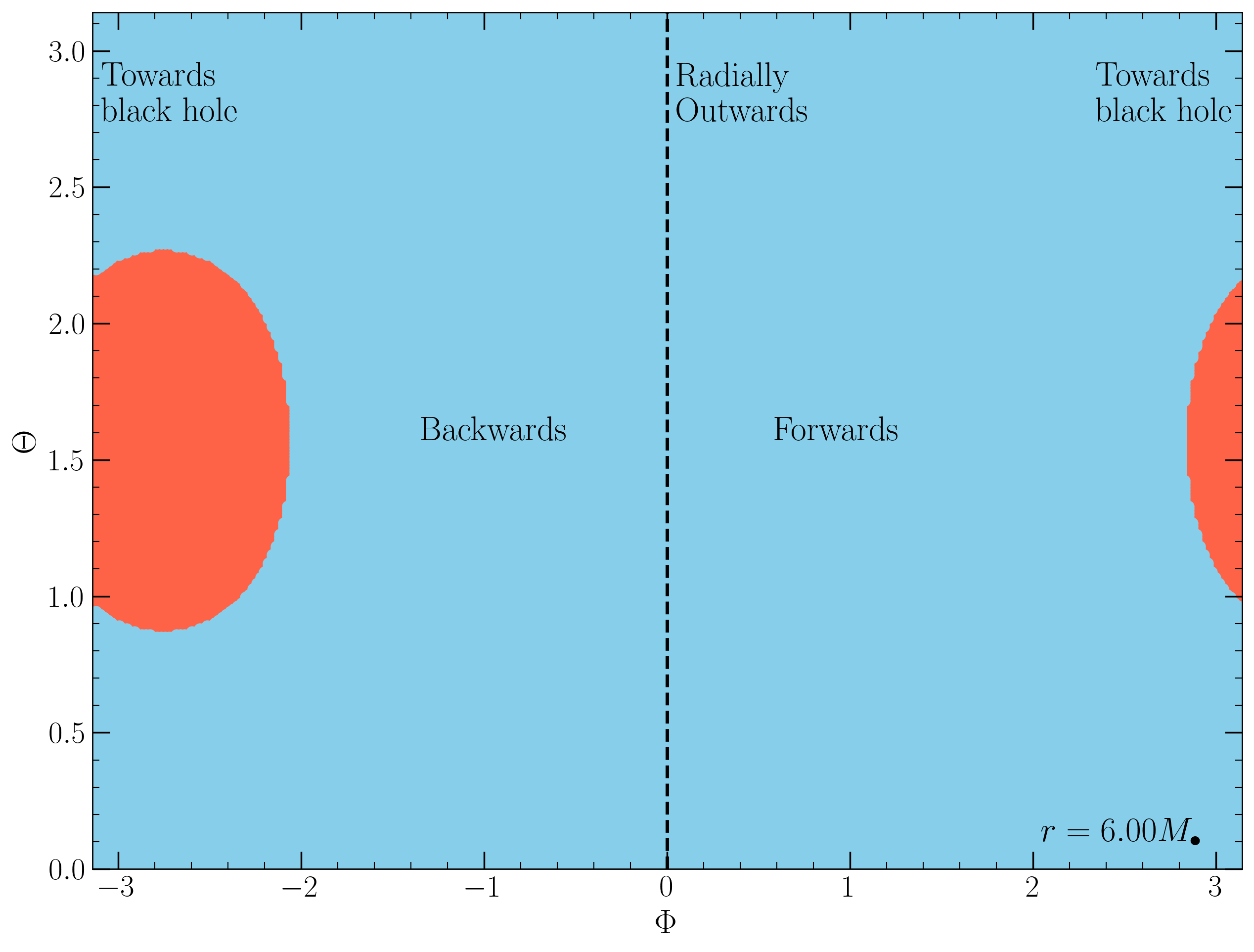}
    \includegraphics[width=0.49\linewidth]{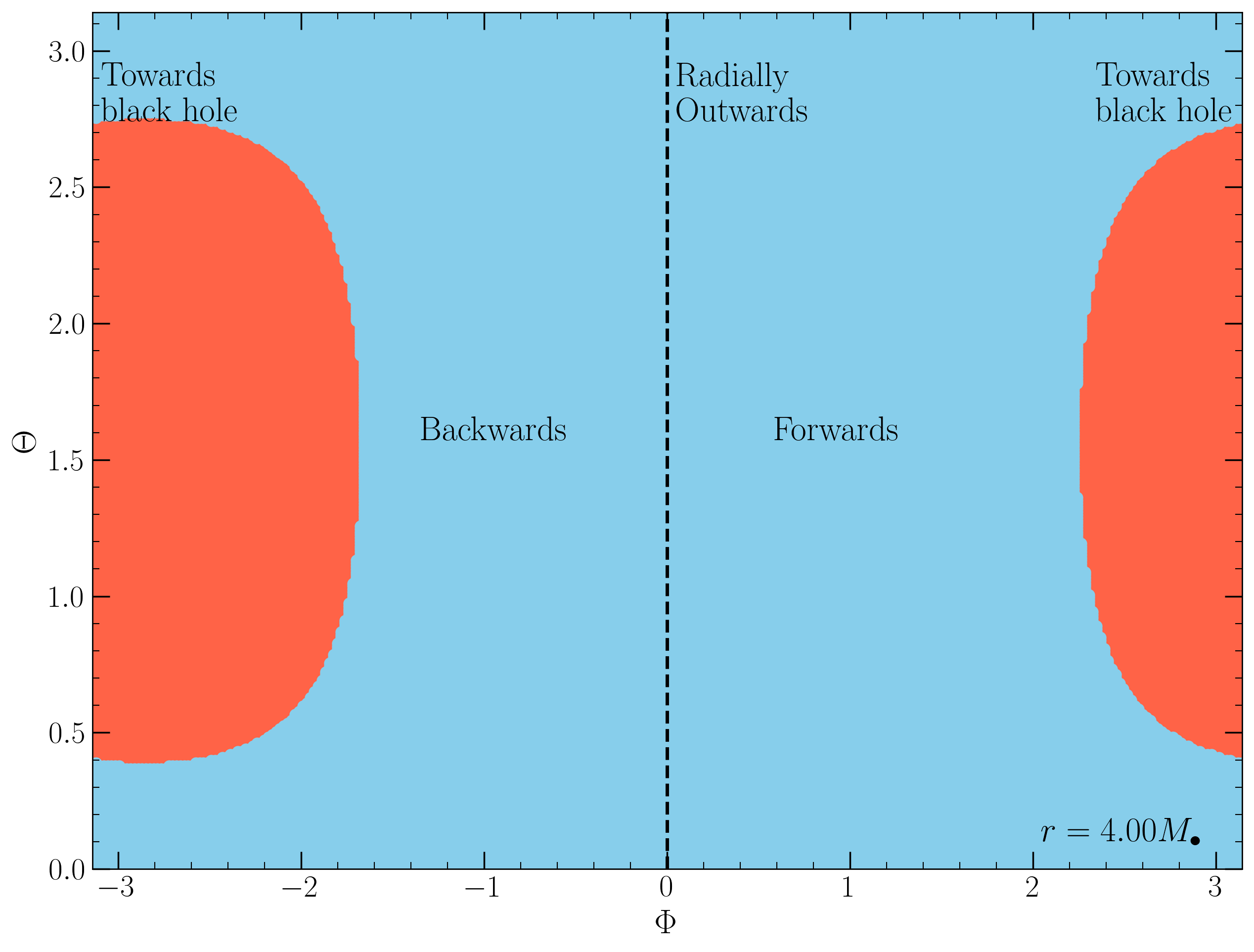}
    \includegraphics[width=0.49\linewidth]{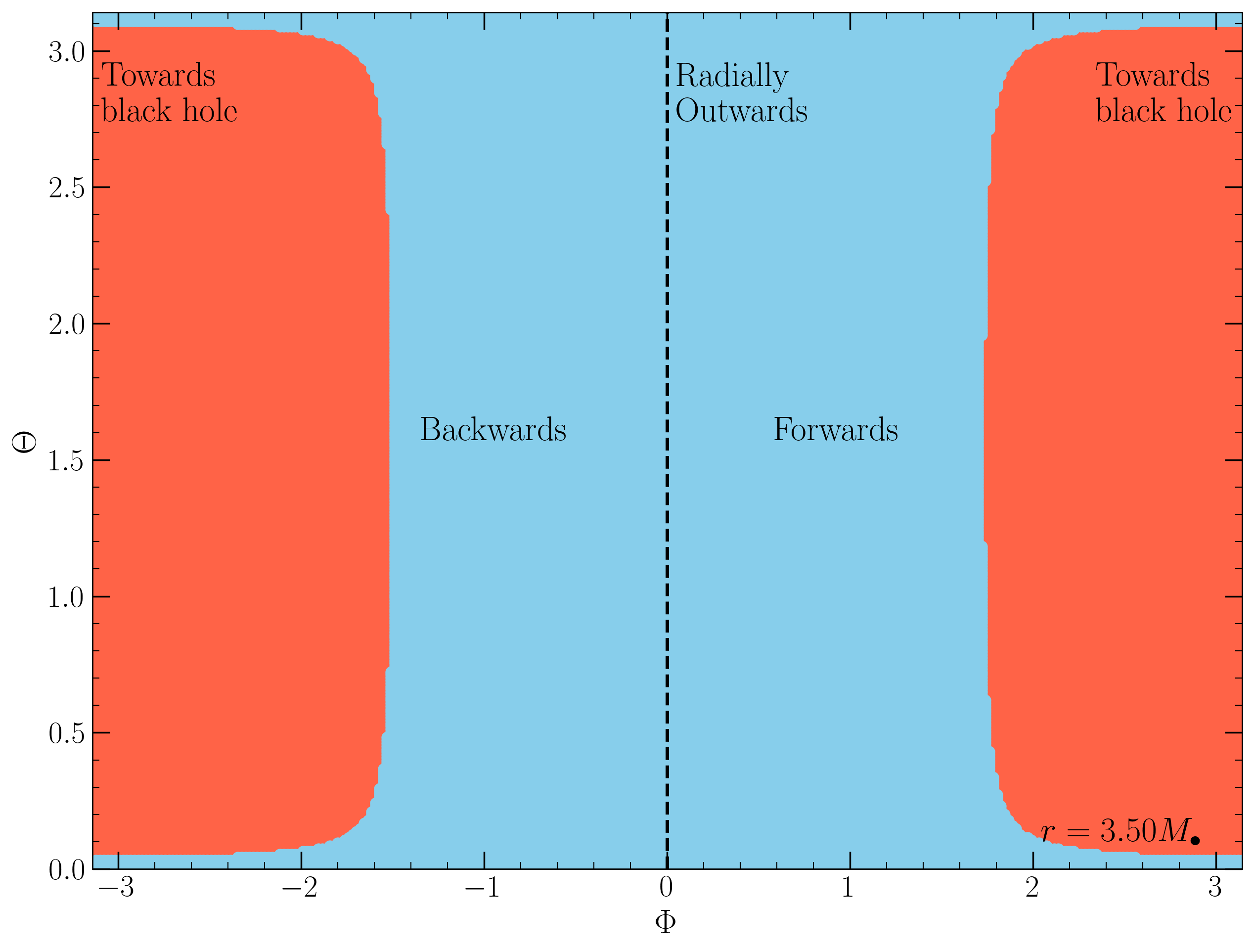}
    \includegraphics[width=0.49\linewidth]{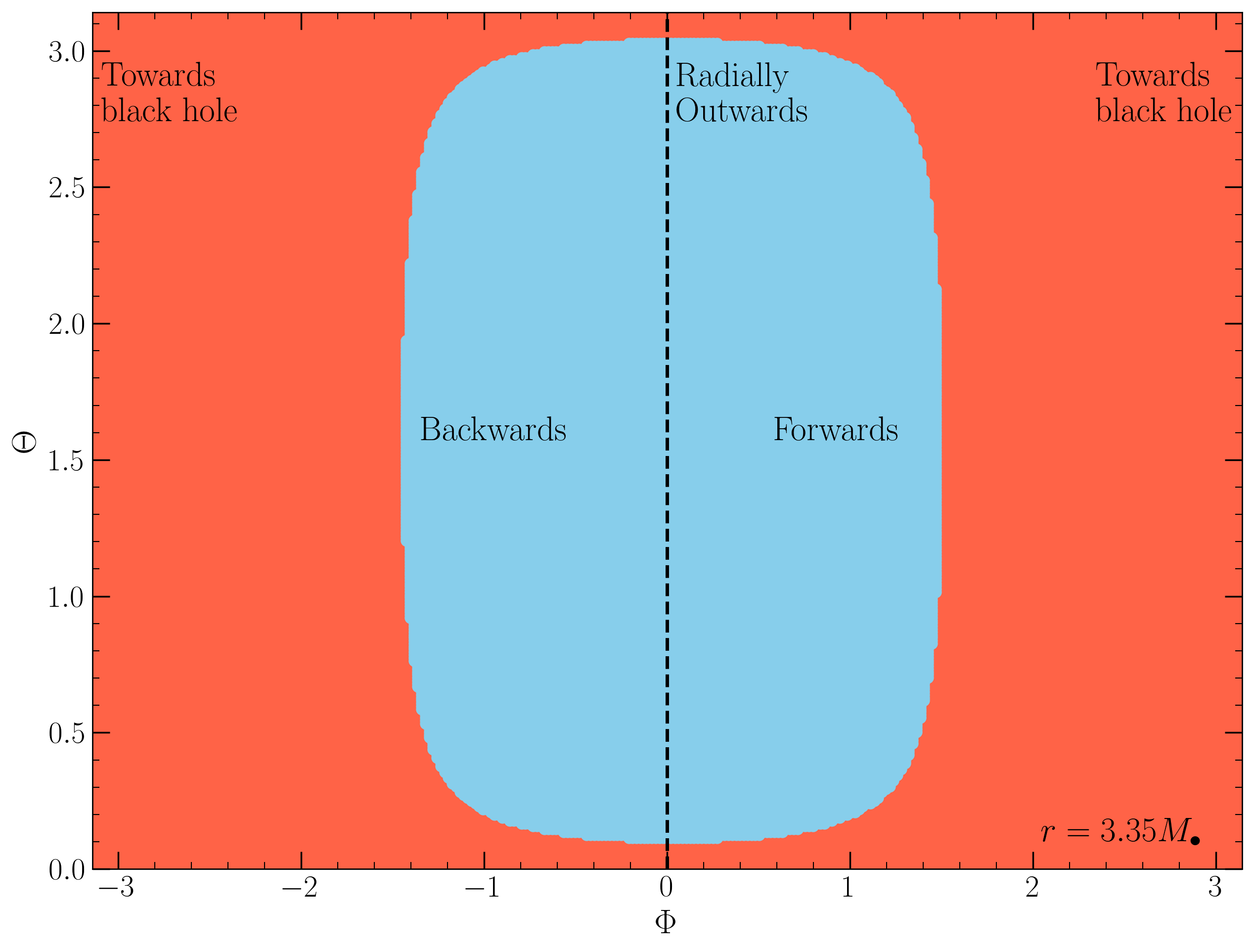}
    \includegraphics[width=0.49\linewidth]{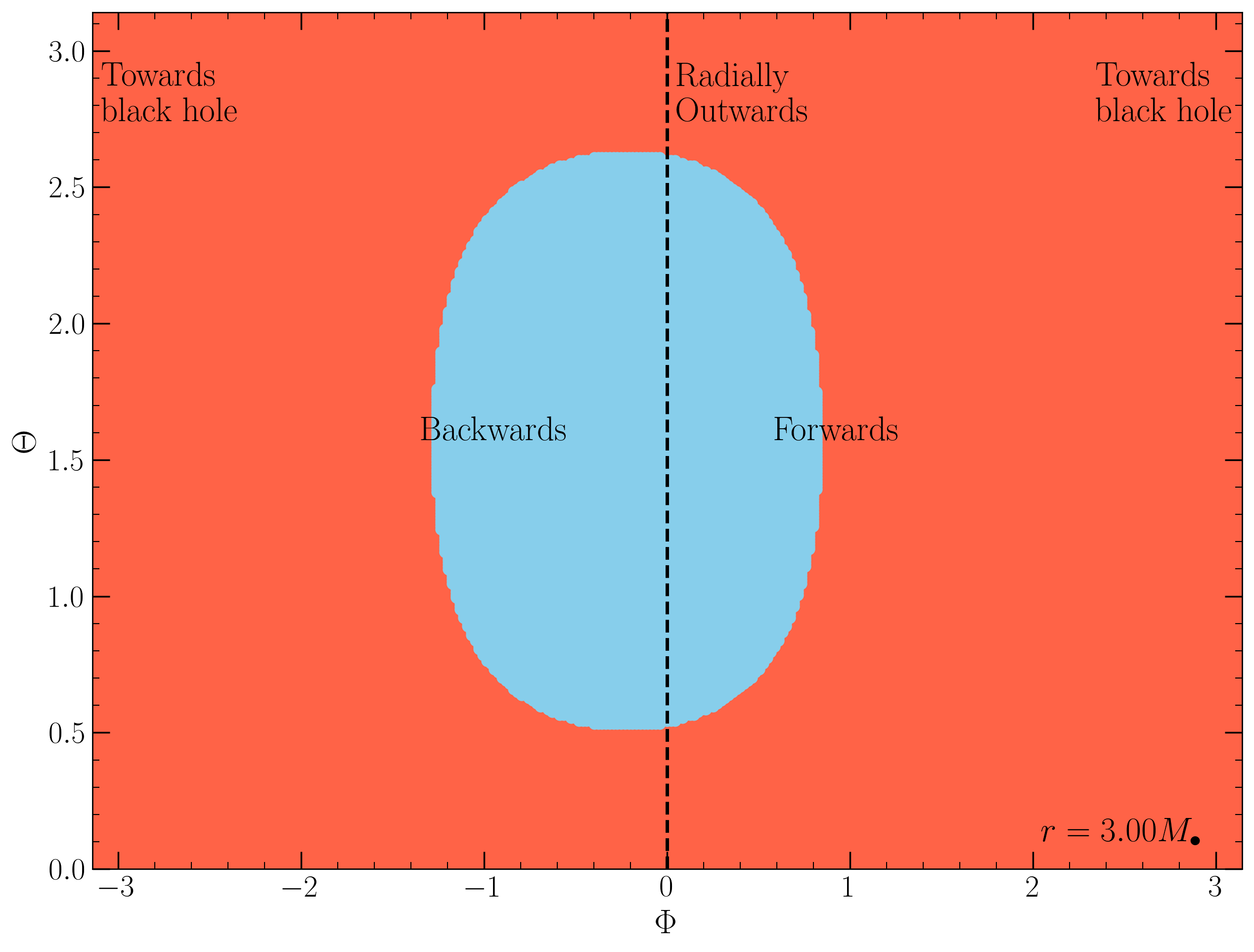}
    \includegraphics[width=0.49\linewidth]{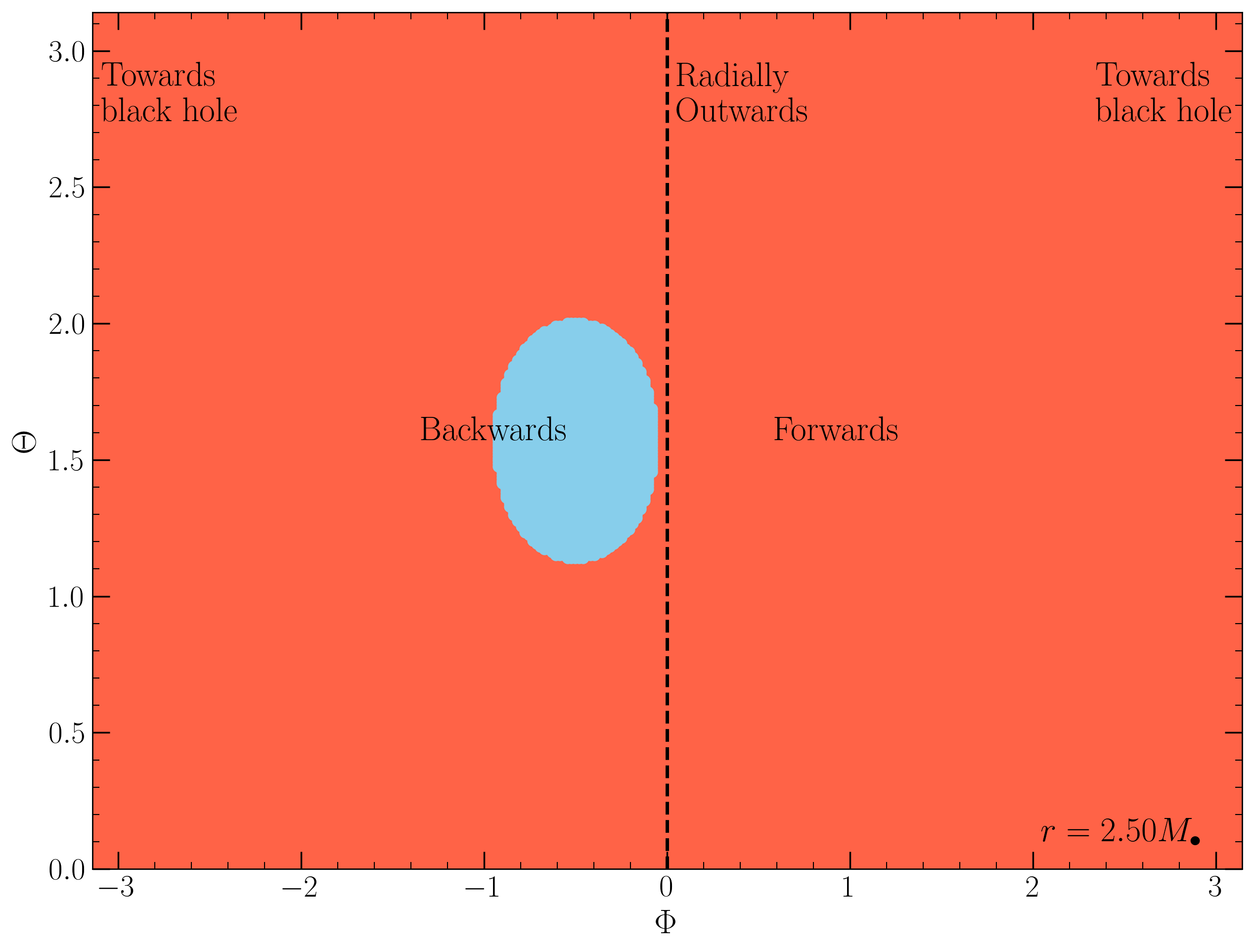}
    \caption{The rest frame angular distribution of the photons captured by (red) and escaping (blue) the black hole, for different emission radii $r$ within the plunging region of a Schwarzschild black hole. The upper left plot shows the fate of photons emitted at the ISCO location, with subsequent panels getting progressively closer to the horizon. The vertical dashed line separates photons emitted (in the fluid rest frame) behind (left of line) and in front (right of line) of the direction of travel of the fluid element. The far left and right of each plot correspond to photons emitted radially towards $r=0$, and the vertical dashed line correspond to photons emitted radially away from the black hole. Each curve is symmetric in the vertical direction about $\Theta = \pi/2$. There is an interesting topological change in the boundary of the captured photon angular distribution at an emission radius $r \simeq 3.4M_\bullet$. Note that the asymmetry in the capture function shifts from being biased towards retrograde photon emission for $r > r_I$ (upper left panel) to being biased towards prograde photon emission for $r \to r_+$ (lower right panel).    }
    \label{fig:pretty}
\end{figure*}

Some photons emitted from the accretion flow are captured by the black hole at all radii. At all radii outside of the ISCO this is typically a relatively small fraction, except for those discs around more rapidly spinning $(a_\bullet \gtrsim 0.9)$ black holes, were the ISCO itself becomes close to the event horizon. Typically for stable disc regions outside of $5M_\bullet$ less than $\sim 10\%$ of all emitted photons are captured, dropping to less than 1\% for $r \gtrsim 10M_\bullet$.

\begin{figure}
    \centering
    \includegraphics[width=\linewidth]{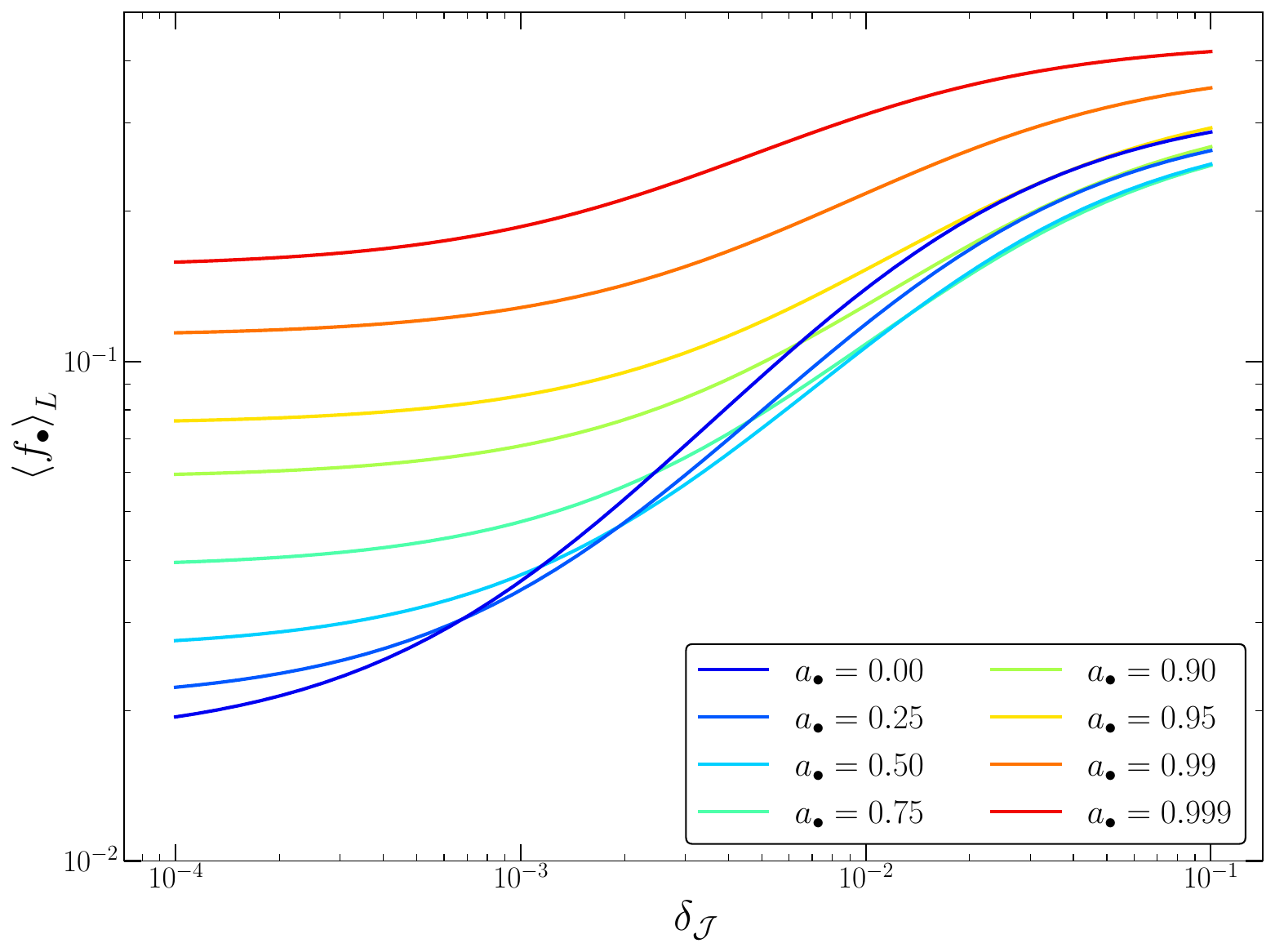}
    \caption{The luminosity-weighted total photon capture fraction, for black hole discs with different spins denoted by line colour (lowest spins have lowest $\left\langle f_\bullet \right\rangle_L$ at lowest ISCO stress) as a function of ISCO stress. At low ISCO stresses the variance in $\left\langle f_\bullet \right\rangle_L$ as a function of black hole spin is large, with higher spins resulting in much larger total photon capture. As the ISCO stress is increased both the total captured fraction increases, but also the variance between different spins decreases, a result of the dropping variance in the radial flux profiles $F(r)$ as more flux is produced, for all spins, at small radii.  }
    \label{fig:fl}
\end{figure}

This picture is radically altered within the plunging region however. Firstly, and obviously, the plunging region extends right down to the event horizon, where $100\%$ of emitted photons will be captured. In addition however, and more interestingly, the fluid's increasing component of radial velocity will result in the relativistic beaming of the emitted radiation in the direction of the event horizon, which naturally results in an increasing capture fraction.  This means that at the same radial scale a photon emitted from plunging material is significantly more likely to be captured than one emitted from stably orbiting material (from a disc around a more rapidly spinning black hole).

To quantitatively examine these effects we compute the local captured fraction $f_\bullet(r)$, which quantifies the fraction of photons emitted at a given radius which ultimately end up crossing the black hole's event horizon. Formally this is given by 
\begin{equation}
    f_\bullet = \int_0^{2\pi} \int_0^{\pi/2}  C(r, \Theta, \Phi) S(\Theta, \Phi) \cos \Theta \sin\Theta  \, {\rm d}\Theta \, {\rm d} \Phi ,
\end{equation}
where in this work we shall consider two emissivity laws 
\begin{align}
    S &= {1\over \pi}, \quad ({\rm isotropic}), \\
    S &= {3\over 7\pi} (1 + 2 \cos \Theta ),  \quad ({\rm electron}\, {\rm scattering}) . 
\end{align}
The first is simple isotropic emission in the rest frame (the factor of $1/\pi$ resulting from the fact that $S$ must be correctly normalised), while the second $S \propto (1 + 2\cos\Theta)$ is that of emission which is “limb-darkened” in the manner expected for an electron-scattering atmosphere. 

The radial structure of $f_\bullet$ is shown in Figure \ref{fig:capfac}. In Fig. \ref{fig:capfac} we show the fraction of emitted photons which are ultimately captured by the black hole  plotted as a function of disc radius for different rest-frame emission laws (top), and for different black hole spins (bottom). For each black hole spin the ISCO radius is denoted by vertical dashed lines, while the horizon radius is denoted by vertical dotted lines. As can be seen for the upper panel comparing different rest-frame emissivity laws, in the innermost $(r \lesssim 5M_\bullet)$ regions there is minimal impact to the capture fraction from different emissivity laws (close to the horizon relativistic radial beaming dominates the capture physics). At larger radii isotropic emission results in more photon capture, this is simply a result of an electron scattering atmosphere sending more photons ``up'' (large $\cos\Theta$) and therefore ``over'' the black hole. {We have verified that our algorithm reproduces the radial dependence of the  photon capture fraction (outside of the ISCO where a comparison can be made) of the works of \cite{Wilkins20b} and \cite{Dauser22}, who each used different algorithms to compute the same results.  }

The rest frame angular distribution of the photons captured by (red) and escaping (blue) the black hole, for different emission radii $r$ of a Schwarzschild black hole ($a_\bullet=0$) is displayed in Figure \ref{fig:pretty}. We have represented the spherical rest-frame emission structure in a two-dimensional $\Theta-\Phi$ plane. So as to be as explicit as possible the vertical dashed line separates photons emitted (in the fluid rest frame) behind (left of line) and in front (right of line) of the direction of travel of the fluid element. The far left and right of each plot correspond to photons emitted radially towards $r=0$, and the vertical dashed line correspond to photons emitted radially away from the black hole. Each curve is symmetric in the vertical direction about $\Theta = \pi/2$.  

We note that there is an interesting topological change in the boundary of the captured photon angular distribution at an emission radius $r \simeq 3.4M_\bullet$, the origin of this effect is that at this radial location the fluid elements are moving at a substantial fraction of the speed of light. This change in topology explains why electron scattering atmospheres (which send more photons with large $\cos \Theta$) have more captured photons within $r \lesssim 3.5M_\bullet$ (Fig. \ref{fig:capfac}). We note that the asymmetry in the capture function shifts from being biased towards retrograde photon emission for $r > r_I$ (upper left panel) to being biased towards prograde photon emission for $r \to r_+$ (lower right panel).  

The effects of black hole spin are more interesting than the effects of differing emissivity shape functions. Higher (prograde) spins push the ISCO closer to the horizon, and therefore prevent relativistic radial-beaming from beginning to dominate the capture physics until much later in the fluid element's evolution. This is because at the same radial scale a photon emitted from plunging material is significantly more likely to be captured than one emitted from stably orbiting material (from a disc around a more rapidly spinning black hole). For each black hole spin the capture fraction approaches $f_\bullet \to 1$ as $r \to r_+$, as it must. For radii $r \gtrsim 10M_\bullet$ there is minimal difference between different black hole spins. 

The {\it total} fraction of emitted photons which are ultimately captured depends on the radial structure of the locally emitted flux $F$, as well as the radial structure of $C$. We define 
\begin{equation}
    \left\langle f_\bullet \right\rangle_L = {\int_{r_+}^\infty f_\bullet(r) \, r F(r) \, {\rm d}r \over \int_{r_+}^\infty  r F(r) \, {\rm d}r} ,
\end{equation}
which is the luminosity-weighted capture fraction, a quantity which  depends only on the black hole spin $a_\bullet$ and the ISCO stress parameter $\delta_{\cal J}$. We plot  the dependence of $\left\langle f_\bullet \right\rangle_L$ on the ISCO stress parameter $\delta_{\cal J}$ in Figure \ref{fig:fl} for a range of black hole spins.

\subsection{Black hole spin evolution}
\begin{figure}
    \centering
    \includegraphics[width=\linewidth]{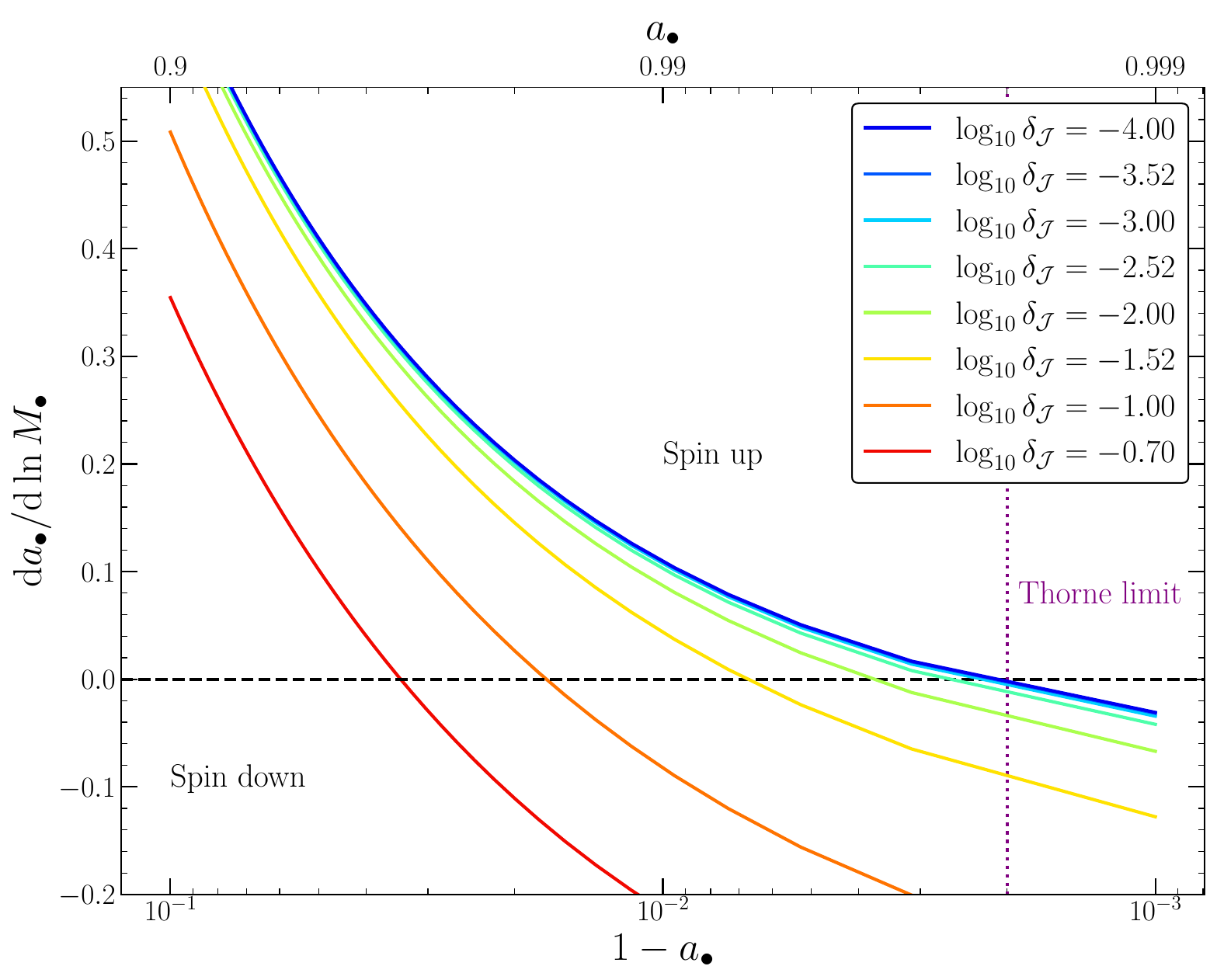}
    \caption{The rate of change of the black hole dimensionless spin parameter during accretion, as a function of spin parameter for different ISCO stress values in the disc. Spin up corresponds to values ${\rm d}a_\bullet / {\rm d}\ln M_\bullet > 0$ (i.e., above the black dashed line), while spin down corresponds to regions below the black dashed line. The equilibrium spin parameter for a given ISCO stress corresponds to the point ${\rm d}a_\bullet / {\rm d}\ln M_\bullet = 0$. The original \citealt{Thorne1974} limit is shown by the vertical dotted purple line. Increasing the ISCO stress parameter results in more emission from within the plunging region, and a lower horizon angular momentum, reducing the black hole spin up rate and resulting in a lower equilibrium spin parameter. GRMHD simulations place the ISCO stress parameter roughly in the range $\delta_{\cal J} \sim 0.02-0.15$, depending on the simulation, reducing the equilibrium spin value to roughly $a_{\bullet, {\rm lim}} \simeq 0.99$.   }
    \label{fig:main_plot}
\end{figure}

Now that we have a full formalism in place, it is possible to examine the effects of an ISCO stress, and increased small length scale emission, on the evolution of the black hole's spin parameter. Our main result is displayed in Figure \ref{fig:main_plot}. 

In Figure \ref{fig:main_plot} we display the value of the total "torque" acting on the black hole as a result of the accretion process, namely  ${\rm d}a_\bullet / {\rm d}\ln M_\bullet$, as a function of black hole spin for a range of ISCO stress parameters. Spin up corresponds to values ${\rm d}a_\bullet / {\rm d}\ln M_\bullet > 0$ (i.e., above the black dashed line), while spin down corresponds to regions below the black dashed line. The equilibrium spin parameter for a given ISCO stress corresponds to the point ${\rm d}a_\bullet / {\rm d}\ln M_\bullet = 0$. We see that for low values of the ISCO stress $\delta_{\cal J} \lesssim 10^{-2}$, the equilibrium black hole spin parameter is largely unchanged from its \cite{Thorne1974} value. This is as expected. 

However, at the level of the ISCO stresses observed in a typical GRMHD simulation $\delta_{\cal J} \sim 0.02-0.15$, the equilibrium black hole spin value begins to deviate  from this value (quite substantially for the more extreme stresses). The dominant effect here is the increased counteracting torque stemming from increased radiative flux over the black hole's horizon. If we introduce the values of the ISCO stress parameter inferred from observations of systems when photons from within the plunging region have been detected \citep[e.g.,][]{Mummery24Plunge, Mummery24PlungeB}, which require $\delta_{\cal J} \simeq 0.04$, then the equilibrium spin is $a_{\bullet, {\rm lim}} \simeq 0.99$. The largest value of $\delta_{\cal J}$ the author is aware of is $\delta_{\cal J} \sim 0.2$, from the simulation of \cite{Noble10}. If this value is accurate, then the equilibrium spin is even lower $a_{\bullet, {\rm lim}} \simeq 0.95$ (red curve Figure \ref{fig:main_plot}). 

As a result of this enhanced retarding radiative torque, the evolutionary histories of black holes accreting via a disc with a finite ISCO stress are modified. This can be most clearly seen by solving the coupled evolutionary equations 
\begin{align}
    {{\rm d}a_\bullet \over {\rm d}m_0} &= {1\over M_\bullet^2}\left[U_\phi(r_+) + {{\rm d}J_{\rm rad}\over {\rm d}m_0}\right] - {2a_\bullet\over M_\bullet}\left[-U_0(r_+) + {{\rm d}E_{\rm rad}\over {\rm d}m_0}\right], \nonumber \\ 
    {{\rm d}M_\bullet \over {\rm d}m_0} &= -U_0(r_+) + {{\rm d}E_{\rm rad}\over {\rm d}m_0} ,
\end{align}
which describe the evolution of the black hole's mass and spin parameters as rest mass (parameterized by $m_0$) is accreted via a disc. 

The solutions of these coupled equations are shown in Figure \ref{fig:evolution}, where we plot the evolution of the black hole mass parameter $M_\bullet$ (upper panel) and spin parameter $a_\bullet$ as rest mass $\Delta m_0$ is added to the hole via thin disc accretion with varying ISCO stresses ($\delta_{\cal J}$). The parameter $M_{\bullet, i}$ represents the initial mass parameter of the hole, which scales out of the problem, and we assume that the initial black hole was described by the Schwarzschild metric (i.e., $a_{\bullet, i} = 0$).  

As an increased ISCO stress parameter results in the black hole capturing more photons (e.g., Figure \ref{fig:fl}), an increased stress results in a more rapid mass evolution (as the photon flux carries a positive energy flux over the horizon), and a slower spin evolution (as on average the photon flux carries a {\it negative} net angular momentum onto the hole). As was highlighted in Figure \ref{fig:main_plot}, this retarding photon torque results in a lower saturation value for the spin parameter for an increasing ISCO stress.  The black dashed line shows a vanishing ISCO stress solution which neglects photon capture (which corresponds to the original Bardeen solution), while the smallest ISCO stress solution is to a very good approximation the original Thorne solution.

\section{Discussion and conclusions}\label{conclusions}

\begin{figure}
    \includegraphics[width=.91\linewidth]{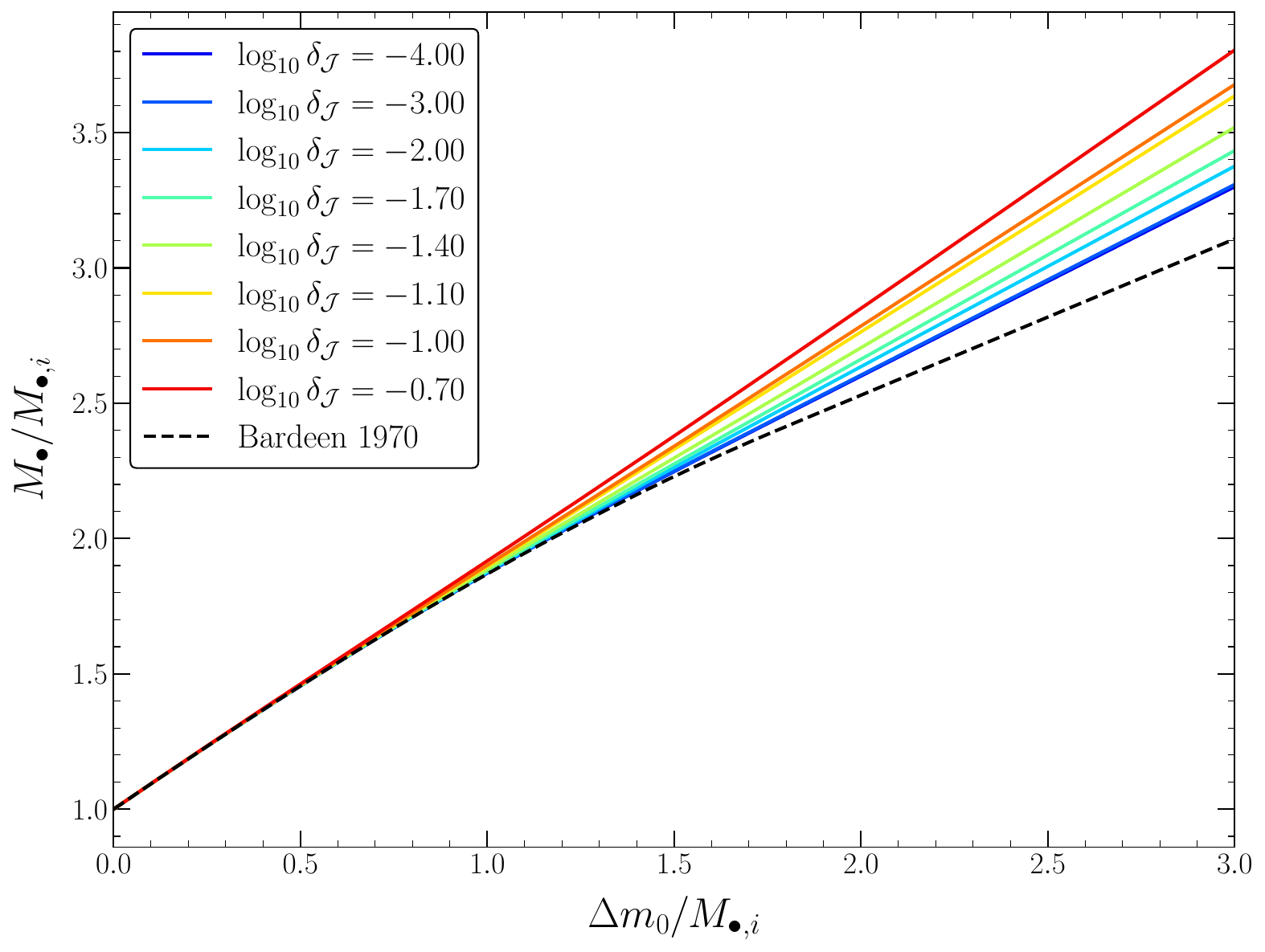}
    \includegraphics[width=1\linewidth]{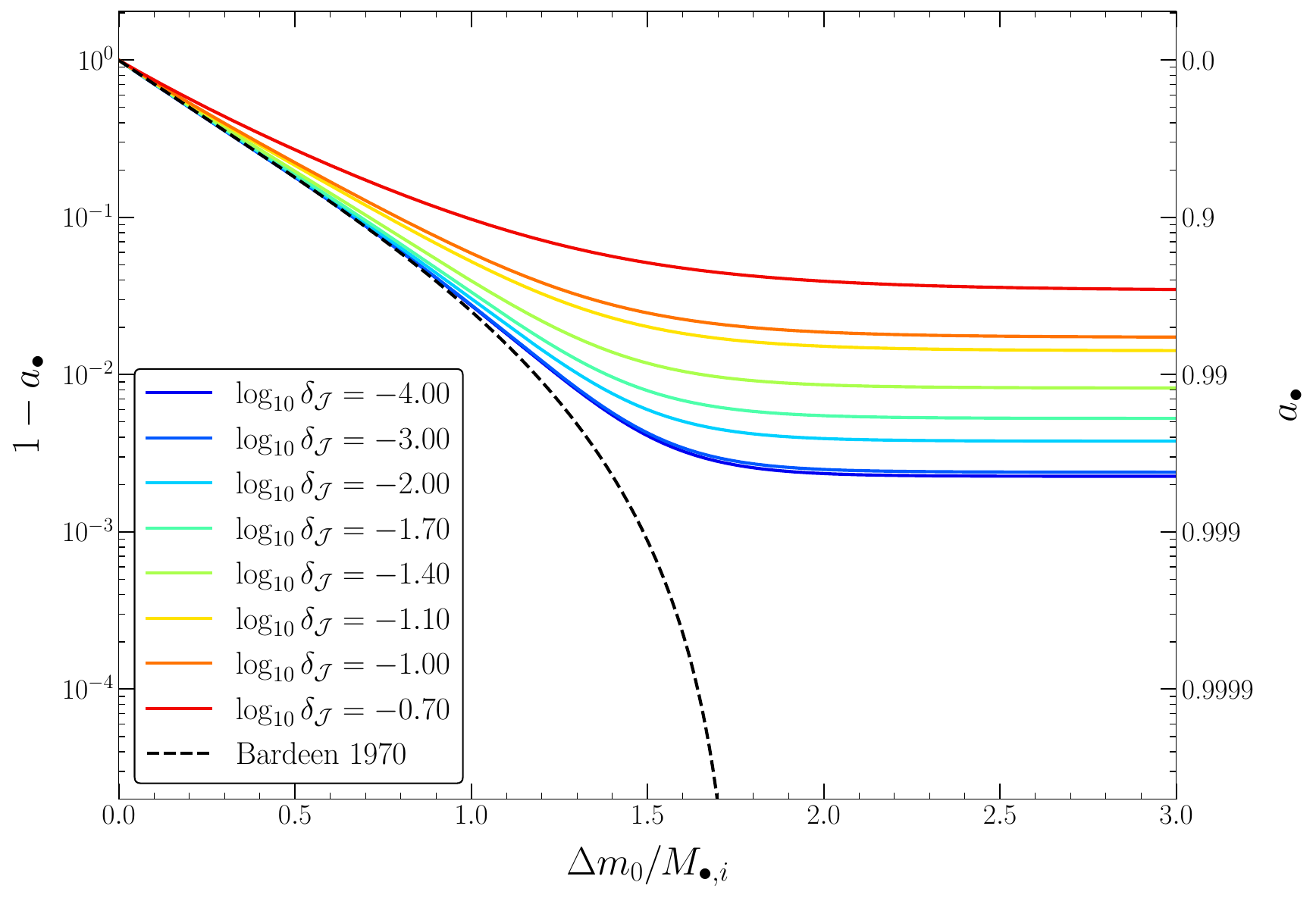}
    \caption{The evolution of the black hole mass parameter $M_\bullet$ (upper panel) and spin parameter $a_\bullet$ as rest mass $\Delta m_0$ is added to the hole via thin disc accretion with varying ISCO stresses ($\delta_{\cal J}$). The parameter $M_{\bullet, i}$ represents the initial mass parameter of the hole. As a result of the hole capturing more photons, increasing the ISCO stress parameter results in quicker mass parameter growth, and retarded spin parameter growth which ultimately saturates at a lower value. The black dashed line shows a vanishing ISCO stress solution which neglects photon capture, which corresponds to the original Bardeen solution.  }
    \label{fig:evolution}
\end{figure}

The main result of this paper is shown in Figures \ref{fig:main_plot} and \ref{fig:evolution}, namely that the finite ISCO stresses which are ubiquitously observed in GRMHD simulations of accretion flows reduce the rate of spin-up of black holes and ultimately the maximum black hole spin parameter which can be reached via thin disc accretion. For this deviation to be non-trivial, the ISCO stress must transport a fraction $\delta_{\cal J} \gtrsim 0.01$ of the ISCO angular momentum back to the disc. The smallest value of $\delta_{\cal J}$ the author is aware of from GRMHD simulations is  $\delta_{\cal J} \simeq 0.02$ \citep{Shafee08}, and observations require $\delta_{\cal J} \simeq 0.04$ \citep{Mummery24Plunge}, which would reduce the spin limit to $a_{\bullet, {\rm lim}} \simeq 0.99$. More recent simulations \citep{Wielgus22} show more substantial deviations $\delta_{\cal J} \simeq 0.1$, which would reduce the spin limit even further. 

To compute this spin limit we have developed techniques describing the photon capture cross sections of emission from within the plunging region, which to the best of our knowledge is the first time such calculations have been performed. Unsurprisingly the fraction of photons captured by the black hole strongly increases once the fluid begins its radial plunge, owing both to the increased proximity of the fluid elements to the horizon, but also to relativistic beaming and a growing radial velocity component. 

While this change in spin parameter is subtle, and it may at first appear that going from the original \cite{Thorne1974} spin-up limit of $a_{\bullet, {\rm lim}} \simeq 0.998$ to $a_{\bullet, {\rm lim}} \simeq 0.99$ is inconsequential, we remind the reader that there are many quantities in the Kerr metric which are extremely sensitive functions of $1 - a_\bullet$ in the limit $1 - a_\bullet\to 0$.  In terms of $1-a_\bullet$, the inclusion of finite magnetic ISCO stress effects results in order of magnitude changes to the final state of the black hole. 

This calculation also highlights that as our models of accretion flows develop and improve from their 1970s forms, it is important to revisit fundamental questions such as the value of the \cite{Thorne1974} limit. There are many ways in which the calculations in this paper could be modified, chiefly in a manner which would further reduce the maximum spin. If black hole jets (often launched during the accretion process) are spin powered \citep[e.g., the][model]{Blandford77} then the presence of jets will inevitably spin down the black hole. Similarly, magnetic field lines which thread both the event horizon and inner edge of the disc can provide a torque on the flow, reducing the maximum spin to a significantly lower value \citep{AgolKrolik00}, namely  $a_{\bullet, {\rm lim}} \simeq 0.36$ (below which point the horizon is rotating more slowly than the inner disc and cannot provide a torque). Finally, and more technically, reheating of the disc by returning radiation will result in even more photon emission from the innermost regions, and reduce the spin limit further from those discussed here. 

We encourage similar calculations of the \cite{Thorne1974} limit to be performed in parallel with developments in accretion theory, as this limiting black hole spin is a quantity of fundamental interest. 

\section*{Acknowledgments} 
I would like to thank Raul for comments on an earlier version of this manuscript. This work was supported by a Leverhulme Trust International Professorship grant [number LIP-202-014]. For the purpose of Open Access, AM has applied a CC BY public copyright license to any Author Accepted Manuscript version arising from this submission. 
 
\section*{Data accessibility statement}
No observational data was taken or used in support of this manuscript. Numerical scripts will be shared upon request to the corresponding author.

\bibliographystyle{mnras}
\bibliography{andy}

\appendix 

\section{Generalised Thorne (1974) photon capture algorithm}\label{algorithm}
The original algorithm presented in \cite{Thorne1974} for determining whether or not an emitted photon is captured by the black hole must be slightly modified for our purposes owing to the non-zero radial velocity of the accretion fluid within the plunging region. So as to be self contained, we reproduce a full version of the algorithm here. As the absolute value of the black hole mass $M_\bullet$ does not effect the calculations in this section, we take $M_\bullet = 1$ without loss of generality.  One begins by picking a pair of rest frame photon emission angles $\Theta$ and $\Phi$, for a photon emitted at radius $r$, and sets the capture function $C(r, \Theta, \Phi) = 1$. One then computes  
\begin{equation}
    n_\mu = g_{\mu \nu} E^\nu_{(a)} \Lambda^{(a)}_{(\alpha)} \widetilde n ^{(\alpha)} ,
\end{equation} 
from which the following quantities related to the photons orbital elements can be defined 
\begin{equation}
    j \equiv a_\bullet^2 - a_\bullet n_\phi /(-n_t), 
\end{equation}
and 
\begin{equation}
    k \equiv (j/a_\bullet)^2 -  n_\theta^2 /n_t^2 . 
\end{equation}
Then the algorithm proceeds as follows, where $r_+ = 1 + (1 - a_\bullet^2)^{1/2}$ is the event horizon radius 
\begin{itemize}
    \item If both $j < -(r_+)^2$ and $r \leq (-j)^{1/2}$ then terminate. 
    \item If both $j < -(r_+)^2$ and $r > (-j)^{1/2}$ then set $C = 0$ and terminate. 
\end{itemize}
If however $j \geq -(r_+)^2$, then one computes the following quantities 
\begin{align}
    \alpha &= 1 + {1\over 3}(j - 2 a_\bullet)^2, \\
    \beta &= 1 - a_\bullet^2 , 
\end{align}
and, if $\beta^2 > \alpha^3$ then
\begin{equation}
    R =1 + (\beta + (\beta^2 - \alpha^3)^{1/2})^{1/3} + \alpha (\beta + (\beta^2 - \alpha^3)^{1/2})^{-1/3}, 
\end{equation}
or otherwise 
\begin{equation}
    R = 1 + 2 \alpha^{1/2} \cos\left({1\over 3} \arccos\left({\beta \over \alpha^{3/2}}\right)\right) .
\end{equation}
Finally, one computes the quantity 
\begin{equation}
    V = {R^2 - 2R + a_\bullet^2\over (R^2 - j)^2} .
\end{equation}
With these quantities calculated, the algorithm proceeds as follows 
\begin{itemize}
    \item If each of $r < R$, $n_r > 0$ and $1/k > V$ are satisfied then set $C = 0$ and terminate. 
    \item If both of $r > R$ and $n_r > 0$ are satisfied then set $C = 0$ and terminate. 
    \item If each of $r > R$, $n_r < 0$ and $1/k < V$ are satisfied then set $C = 0$ and terminate. 
\end{itemize}
If none of these conditions are met, then the algorithm terminates with $C = 1$. The only changes to the \cite{Thorne1974} algorithm used here are in using the radial component of the photons covariant four-velocity $n_r$ in these final steps as opposed to the sign of the emission angle $\cos\Phi$ as used by \cite{Thorne1974}. These algorithms are equivalent for $r \geq r_I$, where the fluid moves on circular orbits. 
\label{lastpage}
\end{document}